\documentclass[twocolumn]{aastex62}	
\usepackage{bm}
\usepackage[colorinlistoftodos]{todonotes}
\hypersetup{colorlinks,citecolor=blue,linkcolor=red}
\usepackage{amsmath}

\accepted{January 13, 2020}
\published{February 14, 2020}
\submitjournal{ApJ}

\shortauthors{Guo et al.}

\begin{document}



\title{Neutrino Production Associated with Late Bumps in Gamma-Ray Bursts and Potential
Contribution to Diffuse Flux at IceCube}




\author[0000-0003-0859-3245]{Gang Guo}
\email{gangg23@gmail.com}
\affiliation{Institute of Physics, Academia Sinica, Taipei, 11529, Taiwan} 

\author{Yong-Zhong Qian}
\email{qianx007@umn.edu}
\affiliation{School of Physics and Astronomy, University of Minnesota, Minneapolis, MN 55455, USA}
\affiliation{Tsung-Dao Lee Institute, Shanghai 200240, China}

\author[0000-0003-4960-8706]{Meng-Ru Wu}
\email{mwu@gate.sinica.edu.tw}
\affiliation{Institute of Physics, Academia Sinica, Taipei, 11529, Taiwan}
\affiliation{Institute of Astronomy and Astrophysics, Academia Sinica, Taipei, 10617, Taiwan}
\affiliation{Physics Division, National Center for Theoretical Sciences, 30013 Hsinchu, Taiwan}


\date{\today}

\begin{abstract} 
IceCube has detected many TeV--PeV neutrinos, but their astrophysical origins remain largely unknown. 
Motivated by the observed late-time X-ray/optical bumps in some gamma-ray bursts (GRBs), 
we examine the correlation between IceCube neutrinos and GRBs allowing delayed neutrinos  
$\sim$days after the prompt gamma rays.
Although we have not found any definitive correlation, up to $\sim$10\% of the events observed so far 
at IceCube may have been neutrinos produced by the late-time GRB activities at $\sim$1 day. Assuming a connection between some IceCube events and the late GRB bumps,
we show in a model-independent way that GRB sites capable of producing 
late $\sim$PeV neutrinos should be nonrelativistic or mildly relativistic.
We estimate the diffuse neutrino flux from such sources and find that they can possibly account for a few IceCube events.
Future observations of high-energy neutrinos and late-time GRB afterglows can further test the above proposed connection.
\end{abstract}   




\section{Introduction}

High-energy (HE) neutrinos of TeV--PeV have been detected by the IceCube Neutrino Observatory \citep{evidence,3years,4years,Aartsen:2017mau}. Recently, an association of neutrino events at IceCube with the blazar TXS 0506+56 has been observed, with the coincidence by chance disfavored at the level of 3$\sigma$--3.5$\sigma$ \citep{Aartsen2018,Ackermann2018}. However, searches for neutrinos from blazars in the third catalog of $Fermi$-LAT sources (3FHL) using the IceCube data indicated that they contribute less than 16.7\% of the diffuse flux at the 90\% confidence level (CL; assuming a spectral index of 2 for blazar neutrinos; \citealt{Huber2019}).
Except for blazars, no other known sources have been found to be correlated with the IceCube events. Therefore, the origin of the majority of these events remains unidentified.

Gamma-ray bursts (GRBs) have long been proposed as one of the most promising sites for producing HE cosmic-rays \citep{Waxman95} and HE neutrinos \citep{waxman1997,waxman1998}.
In the standard fireball model, $\gamma$-rays are produced via synchrotron radiation of electrons accelerated by internal shocks
or via inverse Compton scattering of lower-energy photons on these electrons \citep{piran2005,Meszaros2006,Kumar2014}. Similarly accelerated protons can interact with the $\gamma$-rays to produce charged pions, which then decay to produce HE neutrinos \citep{waxman1997}. These neutrinos are expected to reach the Earth almost simultaneously with the prompt $\gamma$-rays.
However, correlation analyses indicate that these prompt GRB neutrinos can contribute only $\lesssim 1\%$ of the IceCube events
\citep{grb2010,grb2011,grb2014,grb2016,Aartsen2017_grb}.
In the case of long GRBs associated with collapsars, precursor neutrinos may also be expected when the fireball is still propagating inside the stellar envelope that is opaque to $\gamma$-rays \citep{Meszaros2001,Razzaque2003,Razzaque2004,Razzaque2005,Murase2013}.
The contribution of such neutrinos is again tightly constrained by correlation searches with wide time windows \citep{grb2010,grb2011}.

A number of authors \citep{Waxman1999,Dai2000,Dermer2000,Li2002,Murase2007,Razzaque2013,Razzaque2014,Thomas2017}
studied long-term neutrino emission associated with electromagnetic (EM) radiation of the standard GRB afterglow produced by
external shocks. However, the estimated flux of these neutrinos is so low that none should have been detected by IceCube \citep{Razzaque2014}.
Their detection may only be possible with longer exposure time or with the upcoming larger observatories such as IceCube-Gen2 \citep{Gen2} and KM3NeT \citep{KM3NeT}. 

Interestingly, observations show that erratic X-ray/optical flares occur $\sim$$10^{2-3}$ s after the prompt $\gamma$-rays in many long and short GRBs, which cannot be explained by the standard afterglow theory \citep{Zhang:2005}.
HE neutrino flashes could be produced along with these flares and contribute
more to the diffuse neutrino background than the prompt GRB neutrinos \citep{Murase2006}. Detection of HE neutrinos from extended emission, X-ray flares, and plateau emission in short GRBs coincident with gravitational wave signals has also been investigated \citep{Kimura2017}.
In addition, observations show that for a significant fraction of GRBs, late-time X-ray/optical bumps occur with a peak around $t_{\rm p} \sim 1$ day and a width of $\sim$$t_{\rm p}$ \citep{Li2012,Liang2013}. The mechanism producing such late-time emission is unclear, but the associated energy budget can be comparable to or even larger than that of the prompt radiation \citep{Liang2013}. If these bumps are produced by shocks, like the prompt bursts,
then associated production of HE neutrinos can occur, via decay of charged pions from the $p\gamma$ reaction, at $\sim $1 day after the prompt $\gamma$-rays.
 
In this work, we first carry out a search similar to that of \citet{Casey2015} for any correlation between the IceCube events and the prompt emission of GRBs over a time window up to $\pm20$ days. We derive upper limits on the number of IceCube events that can be associated with GRBs for delay times of $\sim$days. Specifically, up to $\sim$10\% of the observed IceCube events may be explained by neutrinos from late-time GRB activities at $\sim$1 day. Assuming the association of late-time EM bumps with HE neutrinos, we further show that strong general constraints on the properties
of the production site can be derived. We estimate the corresponding diffuse HE neutrino flux from the relevant sources and the probability of its detection by IceCube.
We also discuss how future observations can greatly strengthen our  proposed connection between HE neutrinos and late bumps of GRBs.
           
\section{Correlation analysis}
\label{sec:correlation} 

We perform an unbinned likelihood analysis of the correlation
between the IceCube events and the observed GRBs.
The characteristics of the 80 IceCube events observed over six years, including deposited energy, 
observation time, direction (R.A. and Decl.), and the associated errors, are available at \url{http://icecube.wisc.edu/science/data/HE-nu-2010-2014} (the first 53 events) 
and from Table 1 in \citet{Kopper2017} (the last 27 events).
IceCube events caused by atmospheric muons are excluded from our analysis.
GRB samples from 2010 to 2016 May are collected by the Gamma-ray Coordinates Network. We use the samples available from the IceCube collaboration, 
formerly at \url{http://grbweb.icecube.wisc.edu/index.php} (GRBweb1), 
and now at \url{https://icecube.wisc.edu/~grbweb_public/Composition_db.html} (GRBweb2).
For GRBs that are not included or have no direction errors at GRBweb2,
we use the $Fermi$ Gamma-ray Burst Monitor (GBM) Burst Catalog available at \url{https://heasarc.gsfc.nasa.gov/W3Browse/fermi/fermigbrst.html}. Over 1800 GRB samples are included, among which $\sim$85\% are long GRBs, i.e., with durations longer than 2 s.

Following \citet{braun2008} and \citet{Casey2015}, we define the likelihood function as
\begin{align}
{\cal L} = \prod_i^N \Big[ {n_s \over N} S_i + \big(1-{n_s \over N}\big) B_i \Big],
\label{eq:likeli}
\end{align}
where $N=80$ is the total number of HE neutrino events, $n_s$ is the number of neutrinos correlated with GRBs, 
\begin{align}
S_i = {1\over N_{\rm grb}}\sum_j^{N_{\rm grb}} d^S_{ij} \times t^S_{ij}, \;\;\; B_i = {1\over N_{\rm grb}}\sum_j^{N_{\rm grb}} d^B_{ij} \times t^B_{ij}
\end{align}
are the probability density functions (PDFs) for the $i$-th neutrino event under the signal and background hypothesis, respectively,
$N_{\rm grb}=1833$ is the total number of GRBs observed during the six years of concern, $d^{S}_{ij}$ and $t^{S}_{ij}$ are the directional and temporal PDFs for the $i$-th neutrino event and the $j$-th GRB when they are correlated, and $d^{B}_{ij}$ and $t^{B}_{ij}$ are the corresponding PDFs when they are not correlated. Note that for a model-independent study, we only use the directional and temporal information in the analysis.    

For an uncorrelated pair of GRB and neutrino events, both the directional and temporal PDFs are flat. We take $d^B_{ij} = {1\over 4\pi}$ and $t^B_{ij} = {1 \over T_0}$ with $T_0\approx 2200$ days being the total exposure time. As in \citet{grb2016}, the signal directional PDF can be described by 
\begin{align}
d^S_{ij} = {\kappa \over 4\pi\sinh(\kappa)} \exp(\kappa \cos\theta_{ij}), \label{eq:dir}
\end{align}
where $\theta_{ij}$ is the angle between the directions of the $i$-th neutrino event and the $j$-th GRB, and $\kappa=(\sigma_{i}^2+\sigma_j^2)^{-1}$ with $\sigma_{i,j}$ being the direction errors. We add a systematic error $\sigma^{sys}_{grb} = 5.0^\circ$ in the quadratic sum for those GRBs detected by the $Fermi$ GBM \citep{Connaughton2014}. We introduce $\Delta T_{ij} = t^{\nu}_i-t^{grb}_j$ as the observed time difference between the $j$-th GRB and the associated $i$-th neutrino event. Unlike \citet{Casey2015}, we distinguish the two cases in which GRB neutrinos reach the Earth earlier ($\Delta T_{ij}<0$) or later ($\Delta T_{ij}>0$) than the prompt $\gamma$-rays. Assuming that the variation of $\Delta T$ is $\sim\vert \Delta T\vert$, we use a free temporal parameter $T_g$ to specify
\begin{align}   
t^S_{ij}(\Delta T) = \left\{ \begin{array}{llcc} 1/T_g, & \mathrm{if}~T_g<\Delta T<2T_g, \\ 
\quad 0, & \mathrm{otherwise,}  \label{eq:Tg1}\end{array} \right.
\end{align}    
for $\Delta T>0$ ($T_g>0$), and  
\begin{align}   
t^S_{ij}(\Delta T) = \left\{ \begin{array}{llcc} -1/T_g, & \mathrm{if}~2T_g<\Delta T<T_g, \\ 
\quad 0, & \mathrm{otherwise,}  \label{eq:Tg2}\end{array} \right.
\end{align} 
for $\Delta T<0$ ($T_g<0$).

We define the test statistic as
\begin{align}
\lambda_{T_g} = 2 \ln[ {\cal L}(\hat n_s)/{\cal L}(n_s=0)],
\end{align}
where $\hat n_s$ is the best-fit value of $n_s$ at which ${\cal L}$ reaches its maximum value for a given $T_g$. Below we follow \citet{grb2014} and \citet{Casey2015} to calculate the $p$-values and the upper limits. 

\begin{table*}[htbp] 
\centering 
\renewcommand{\arraystretch}{1.5}
\begin{tabular}{p{1.cm}<{\centering}p{1.1cm}<{\centering}p{1.cm}<{\centering}p{1.cm}<{\centering}p{1.cm}<{\centering}|p{1.8cm}<{\centering}p{1.cm}<{\centering}p{1.cm}<{\centering}p{1.cm}<{\centering}p{1.cm}<{\centering}|p{1.cm}<{\centering}p{1.cm}<{\centering}}    
\hline
\hline
\multicolumn{5}{c|}{IceCube HESEs} & \multicolumn{5}{c|}{GRBs} & \multicolumn{2}{c}{} \\     
\hline
ID & $E_{\rm dep}$ (TeV) & Decl.\ \ ($^\circ$) & R.A.\ \ \ ($^\circ$) & Error\ \ ($^\circ$) & GRB No. & Decl.\ \ ($^\circ$) & R.A.\ \ \ ($^\circ$) & Error\ \ ($^\circ$) & Long/ Short & $t^\nu_i-t^{\rm grb}_j$ (day) & $\theta_{ij}$\ \ \ \ \ \ ($^\circ$) \\
\hline
63 & 97.4 & 6.5 & 160.0 & 1.2 & GRB141207A & 3.7 & 159.9 & $10^{-3}$ & L & 1.34 & 2.8   \\ 
50 & 22.2 & 59.3 & 168.6 & 8.2 & GRB140320B & 60.3 & 145.6 & 0.05 & L & 0.81 & 11.6 \\
14 & 1041 & -27.9 & 265.6 & 13.2 & GRB110808B & -37.7 & 266.2 & 0.07 & S & 0.87 & 9.8 \\
9 & 63.2 & 33.6 & 151.3 & 16.5 & GRB110503A & 52.2 & 132.8 & $10^{-4}$ & L & 0.93 & 22.9 \\ 
23 & 82.2 & -13.2 & 208.7 & 1.9 & GRB120121C  & -1.34 & 208.9 & 5.3 & L & 2.3 & 11.9 \\ 
\hline
\end{tabular}
\caption{Potentially correlated pairs of IceCube HESEs and GRBs with the largest values of $d^S_{ij} \times t^S_{ij}$ for $T_g>0.1$ day. The errors of the IceCube events and the GRBs are the median angular errors \citep{evidence,3years,4years} and the 1$\sigma$ angular errors assuming a 2D Gaussian distribution, respectively.}
\label{tab:pair}
\end{table*} 

\begin{figure}
\centering
\includegraphics[width=8.5cm]{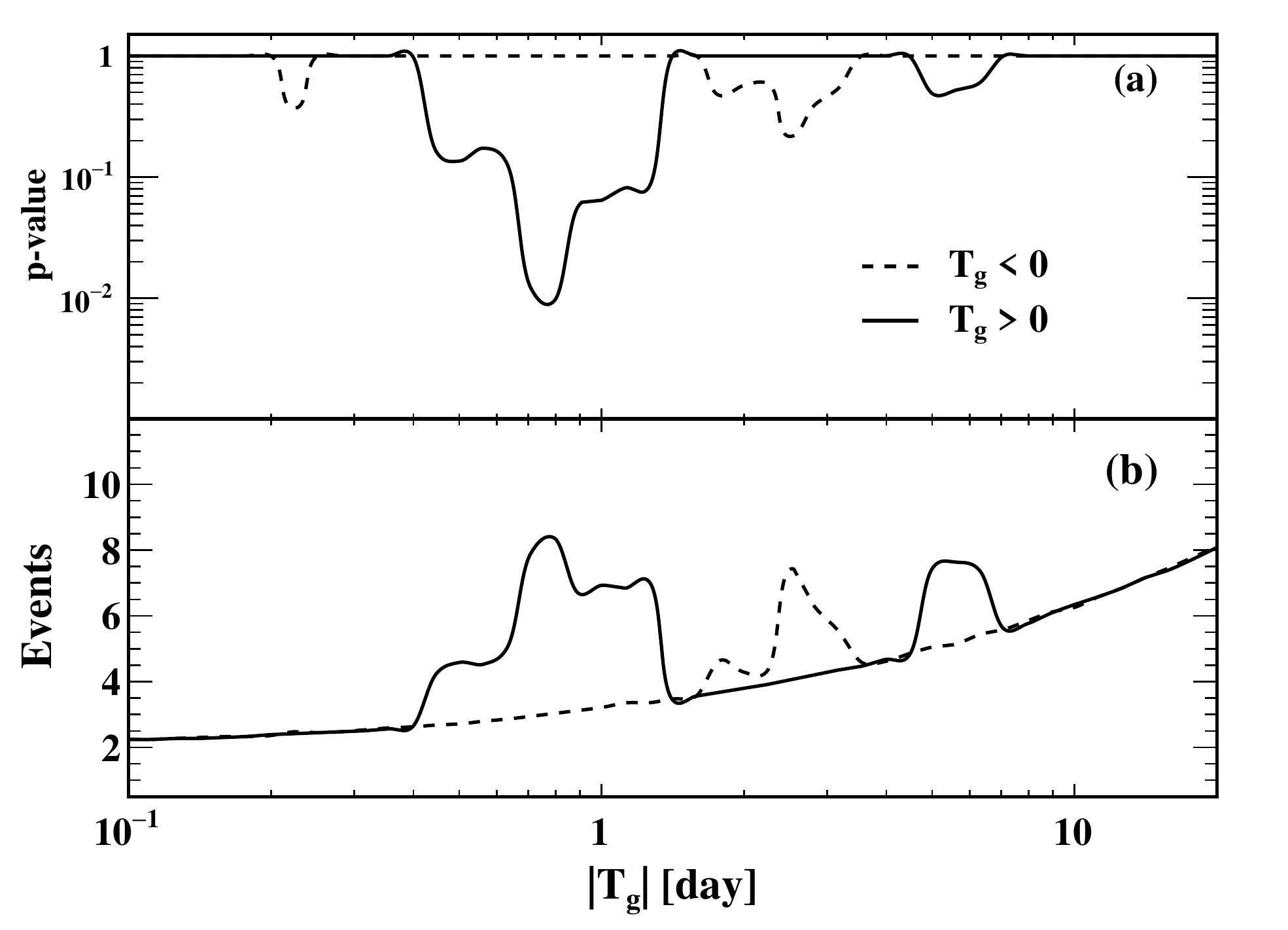} 
\caption{Panel (a): $p$-value as a function of $T_g$, which characterizes the time difference between a correlated pair of neutrino event and GRB.
Panel (b): 90\% CL upper limit on the number of neutrino events correlated with GRBs as a function of $T_g$.}
\label{fig:prep-Up}
\end{figure}

We carry out $4\times 10^4$ simulations of the data set by randomizing the directions and arrival times of all the relevant IceCube events, while keeping their directional errors unchanged. We obtain a distribution of $\lambda_{T_g}$ based on these simulated data sets. The $p$-value for a given $T_g$ is the probability of finding $\lambda_{T_g} > \lambda_{T_g}^{\rm obs}$ in the distribution, where $\lambda_{T_g}^{\rm obs}$ is the test statistic based on the true data set. The lower the $p$-value is, the more likely there is a true correlation. Figure \ref{fig:prep-Up}(a) shows the $p$-value as a function of $T_g$. We find that the most significant $p$-value, i.e., the pre-trial $p$-value, $p_{\rm pre} \approx 9.5\times 10^{-3}$, occurs at $T_g \approx 0.78$ day with $\hat n_s \approx$ 4.7. To account for the trial factor, we take each simulated data set as if it is the true one and follow the procedure above to obtain a distribution of the pre-trial $p$-values. The post-trial $p$-value is the probability of finding a $p$-value in this distribution that is more significant than the pre-trial $p$-value for the observed data. We find $p_{\rm post} \approx 0.47$, indicating that the data are consistent with the null hypothesis. 

Consequently, we can set an upper limit on the number of IceCube events correlated with GRBs as a function of $T_g$. To obtain the upper limit, we need to simulate data sets including different numbers of signal events. We randomly choose $N_i$ IceCube events and pair each with a randomly chosen GRB to simulate the signal events. We then randomly generate the directions and arrival times of the selected IceCube events using the probability distributions in Equation~(\ref{eq:dir}) and Equation~(\ref{eq:Tg1}) or (\ref{eq:Tg2}) for a given $T_g$. The remaining $80-N_i$ neutrino events are simulated as background. For calculating the upper limit, we choose $N_i$ to be 0, 1, 2, ..., 30, and $T_g=\pm 10^{0.05 i-1}$ day with $i=0$, 1, ..., 46. For each given $T_g$ and $N_i$, we simulate $10^4$ data sets and obtain a normalized distribution $P_{N_i}(\lambda_{T_g})$ of the test statistic $\lambda_{T_g}$. For a given $T_g$ and a given mean value $\langle n_s \rangle$ for the number of correlated events, we define a distribution $P_{\langle n_s \rangle}(\lambda_{T_g})=\sum_{N_i} P_{N_i}(\lambda_{T_g})
\exp(-\langle n_s \rangle) \langle n_s \rangle^{N_i}/N_i !$. The 90\% CL upper limit $\langle n_s \rangle^{\rm up}$ for a given $T_g$ corresponds to $\int_{\lambda_{T_g}^{\rm obs}}^\infty P_{\langle n_s \rangle^{\rm up}}(\lambda_{T_g})d\lambda_{T_g}= 0.9$, where $\lambda_{T_g}^{\rm obs}$ is for the observed data. The 90\% CL upper limit $\langle n_s \rangle^{\rm up}$ is shown as a function of $T_g$ in Figure \ref{fig:prep-Up}(b). The allowed correlated event number tends to increase with $|T_g|$ simply due to random coincidence. However, excesses above this smooth general
trend may indicate true correlation. In particular, the large excess at $T_g \sim 1$ day indicates that up to $\sim$10\% of the events observed so far at IceCube might have been produced by late-time GRB activities on this time scale, which
motivates us to further explore the possible connection between HE neutrinos and late GRB bumps. Note that our results are in quantitative agreement with those in \citet{Casey2015} except that we explicitly distinguish the cases of $T_g>0$ and $T_g<0$. 

A few possibly correlated pairs with the largest values of $d_{ij}^S \times t_{ij}^S$ for $T_g>0.1$ day are listed in Table~\ref{tab:pair}. Unfortunately, no long-term optical data at $\sim$1 day were recorded for any possibly correlated GRBs. For GRB110503A, X-ray afterglow was observed up to $10^6$ s but no bump was seen. So current observations are unable to shed light on the association of HE neutrinos with the late EM bumps. Nevertheless, we highlight a few possibly correlated observations. A $1.04$~PeV IceCube shower event is potentially correlated with the very intense and short-hard GRB110808B. Their reported directions are within $1\sigma$ error, and the arrival time of the neutrino event is $\sim$21 hr after the GRB. In addition, the most energetic track event with $E_\mathrm{dep} = 2.6$ PeV (not included in our analysis) observed recently \citep{HE1,HE2} may be correlated with GRB140610C, arriving $\sim$16 hr after this bright long burst, which has a systematic error of $4^\circ$--$10^\circ$ in its direction \citep{Connaughton2014}. Assuming the connection between HE neutrinos and late GRB bumps, below we study the implications for the sites capable of producing these neutrinos.

\section{Model-independent constraints}

The late flares or bumps are believed to be related to the late central engine activities of GRBs \citep{Burrows:2005,Zhang:2005}. Although the exact origin remains unclear, various mechanisms were studied, including scenarios with two-component jets \citep{Berger2003}, refreshed shocks \citep{Rees1997,Kumar1999,Sari2000}, late reverse shocks \citep{Kobayashi2003,Zhang2003}, and density bumps \citep{Lazzati2002,Dai2003}, etc. Motivated by these studies and the hint from our analysis of the correlation between IceCube events and GRBs, we consider that HE neutrinos are produced by $p\gamma$ reactions between HE protons and photons of the late bump, both of which arise from particle acceleration by shocks in some late outflows. We show that stringent constraints can be put on the relevant site, which is required to accelerate protons to sufficient energy, facilitate efficient transfer of energy from protons to neutrinos, and promote HE neutrino production by avoiding meson cooling.
 
\subsection{Accelerating protons}
As cooling due to synchrotron radiation and the Bethe-Heitler process ($p\gamma\to pe^-e^+$) is less significant for the energy range explored, the maximal energy of accelerated protons, $E'^{\rm max}_p$, can be estimated by equating the cooling time scale due to $p\gamma$ reactions, $t'_{p\gamma}$, and the time scale for proton acceleration by shocks, $t'_{\rm acc}$. Here and below, primed quantities refer to the comoving frame of the shocked outflow. 

The photon flux of the bump from the optical to X-rays is observed to follow a simple power law  $dn(\epsilon)/d\epsilon \sim \epsilon^{-\Gamma_\gamma}$ with $\Gamma_\gamma\sim 1.5$--2 \citep[see, e.g.,][]{Margutti2010,Melandri:2014}.
We assume that this form extends from 0.1 eV to some photon energy $E_c$, which is taken to be 100 keV. 
The lower bound is irrelevant, as photons with energy lower than 0.1 eV are below the threshold of $p\gamma$ reactions and do not contribute to production of HE neutrinos of $\lesssim$ 10 PeV. 
We have also checked that our results are affected very little when varying $E_c$ from 10 keV to 1 MeV.
We take $\Gamma_\gamma=2$ to better fit the observed X-ray luminosity, which is typically $\sim$10 times higher than the $R$-band luminosity for the bump \citep{Li2012,Liang2013}. The normalization of the photon spectral density is determined by the optical $R$-band (520--800 nm) isotropic luminosity of the bump
\begin{align}
L^{\rm iso}_{\rm R} = \int_{R\mbox{-}\rm{band}} 4\pi R^2c\epsilon \frac{dn(\epsilon)}{d\epsilon} d\epsilon,
\end{align}
where $R$ is the typical shock radius. Note that
$L^{\rm iso}_{\rm R}$ is measured in the stellar rest frame, and has typical values of $10^{45}$ erg/s at $t_{\rm ob} \sim$ 1 day \citep{Kann2010,Li2012,Zaninoni2013}.
Relating the comoving frame to the stellar rest frame, we have
$dn^\prime(\epsilon^\prime)/d\epsilon^\prime \approx dn(\epsilon)/d\epsilon$, where $\epsilon \approx \Gamma \epsilon'$ is the photon energy in the latter frame and $\Gamma$ is the Lorentz factor of the shocked fluid.

Taking the $\Delta$-resonance approximation \citep{Murase2016}, we estimate 
\begin{align}
t'^{-1}_{p\gamma} & \approx \hat\sigma_{p\gamma} c (\epsilon^\prime dn^\prime/d\epsilon^\prime)]\vert_{\epsilon^\prime = 0.5m_pc^2\bar\epsilon_\Delta/E_p^\prime} \nonumber \\ 
&\approx 4.9 \times 10^{-15} ~(R^2_{17}\Gamma^2)^{-1} L_{\rm R, 45}^{\rm iso} (E'_p/{\rm GeV})~{\rm s}^{-1},   
\label{eqn:tpg}
\end{align}
where $\hat \sigma_{p\gamma} \approx 0.6 \times 10^{-28}~\rm{cm^2}$ is the $p\gamma$ cross section in the resonance limit,\footnote{The value of $\hat\sigma_{p\gamma}$ corresponding to the $\Delta$-resonance approximation is about $1.5 \times 10^{-28}~{\rm cm^2}/(1+\Gamma_\gamma) = 0.5 \times 10^{-28}~\rm cm^2$. A slightly larger $\hat\sigma_{p\gamma}$ is used in Equation~(\ref{eqn:tpg}) to match the result from the full calculation presented in Appendix~\ref{sec:pr}.} $\bar\epsilon_\Delta \approx 0.3$ GeV, and the notation $A_{x}$ means $A/10^x$ in cgs units. The analytical expression in Equation~(\ref{eqn:tpg}) agrees pretty well with the detailed calculation of $t'_{p\gamma}$ based on a more accurate $p\gamma$ cross section (see Appendix~\ref{sec:pr}).   

The acceleration time scale is approximately given by    
\begin{align}
t'_{\rm acc} \sim \theta_{\rm F} E_p'/(eB'c) \sim 10^{-3}~(E'_p/{\rm GeV})({\rm G}/B')~{\rm s}, 
\end{align}
where $\theta_{\rm F}=10$ is the acceleration constant used in our study \citep{Rachen:1998}, and $B'$ is the magnetic field. 
In the literature, $\epsilon_B$ and $\epsilon_e$ are usually introduced as the fractions of the internal energy of the shocked fluid transferred to the magnetic field and electrons, respectively. All the energy of electrons is emitted in EM radiation. From the observed photon flux, we can then estimate the total kinetic energy of the shocked outflow and the energy carried by the magnetic field. The energy density of electrons is given by $U'_e = L_\gamma^{\rm iso}/(4 \pi R^2 \Gamma^2 c) = L^{\rm iso}_{\rm R}/(4 \pi \epsilon_{\rm R} R^2 \Gamma^2 c)$, where $\epsilon_{\rm R} = L^{\rm iso}_{\rm R}/L^{\rm iso}_\gamma$ is the ratio of the photon energy in the $R$ band to that in all bands. With $U'_B = B'^2/(8\pi) = U'_e \epsilon_B/\epsilon_e$, the magnetic field is 
\begin{align}
B' =& \sqrt{\frac{2L^{\rm iso}_{\rm R} \epsilon_B}{\epsilon_{\rm R} \epsilon_e R^2 \Gamma^2 c}} \nonumber \\
\approx & 15 \sqrt{L_{\rm R, 45}^{\rm iso} R_{17}^{-2} \Gamma^{-2} \Big(\frac{\epsilon_B}{0.1}\Big) \Big(\frac{0.3}{\epsilon_{\rm R}}\Big) \Big(\frac{0.1}{\epsilon_e}\Big)}~{\rm G}.         
\end{align}

With $t'_{\rm acc}=t'_{p\gamma}$ at $E_p'=E'^{\rm max}_p$, producing a typical HE neutrino of $E_{\nu, {\rm ob}} \approx 0.05E'_p\Gamma/(1+z)$ requires
\begin{align}
R_{17} \Gamma^3 \ge 5\times 10^{-4}~ \Big(\frac{E_{\nu,{\rm ob}} \hat z}{\rm PeV}\Big)^2 \Big( L^{\rm iso}_{\rm R, 45} \frac{\epsilon_{\rm R}}{0.03} \cdot \frac{\epsilon_e}{0.1}\cdot \frac{0.1}{\epsilon_B} \Big)^{1/2},  \label{eq:acc}   
\end{align}  
where $\hat z\equiv(1+z)/2$ with $z$ being the redshift. 

\subsection{Energy transfer by $p\gamma$ reactions}

The fraction of the energy transferred from protons to neutrinos can be estimated as   
\begin{align}
f_{p\gamma} \sim t'_{\rm dyn}/t'_{p\gamma} 
&\sim 1.63\times 10^{-8} ~(R_{17}\Gamma^3)^{-1} L_{\rm R, 45}^{\rm iso} (E'_p/{\rm GeV})\nonumber\\
&\sim 0.65~ (R_{17}\Gamma^4)^{-1} L^{\rm iso}_{\rm R, 45} (E_{\nu,\rm ob}\hat z/\rm PeV), \label{eq:fpr0}
\end{align}  
where $t^\prime_{\rm dyn}\sim R/(\Gamma c)$ is the dynamical time scale, and we have used
$E_{\nu, {\rm ob}} \approx 0.05E'_p\Gamma/(1+z)$ in the second line.
Hence, efficient transfer requires 
\begin{align}
 R_{17} \Gamma^4 \lesssim 0.65~ L_{\rm R, 45}^{\rm iso} \frac{E_{\nu, {\rm ob}}}{\rm PeV}\hat z. \label{eq:fpr}
\end{align}

\begin{figure}
\centering
\includegraphics[width=7.5cm]{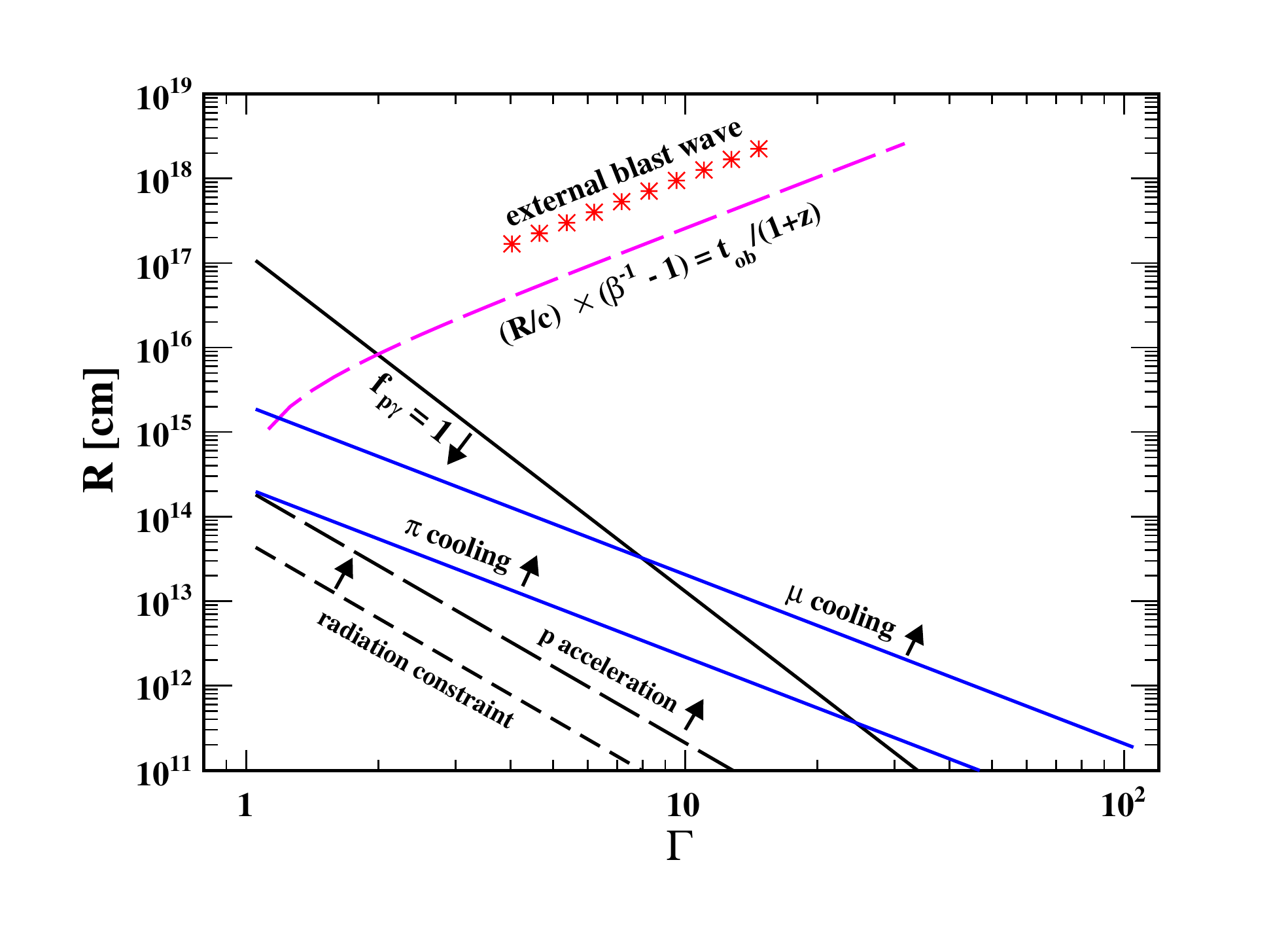}%
\caption{Bounds on the shocked outflow radius $R$ and Lorentz factor $\Gamma$ for efficient production of PeV neutrinos from late GRB bumps.
The red crosses represent typical $R$ and $\Gamma$ for the standard GRB blast wave at $t_{\rm ob}=1$ day. See the text for details.} 
\label{fig:RGamma}
\end{figure} 

\subsection{Avoiding meson cooling} 
Because production of HE neutrinos relies on the decay of $\pi^\pm$ and $\mu^\pm$, the latter particles should not suffer significant energy loss from synchrotron cooling. For charged particles (including protons, $\pi^\pm$, and $\mu^\pm$) with mass $m_i$ and energy $E'_i$, the synchrotron cooling time scale is given by 
\begin{align}
t'_{{\rm syn}, i} &= \frac{6\pi m_i^4 c^3}{ \sigma_{\rm T} \beta'^2_i m_e^2 E'_i B'^2} \nonumber \\
& \approx 2.6 \times 10^{16}~\Big(\frac{m_i}{\rm GeV}\Big)^4 \Big(\frac{\rm GeV}{E'_i}\Big) \Big( \frac{15~\rm G}{B'}\Big)^2~{\rm s}, 
\end{align}
where $\sigma_{\rm T} \approx 6.65 \times 10^{-25}~\rm{cm^2}$ is the Thomson cross section, and $\beta'_i = v_i/c \approx 1$ is the velocity of the charged particle in units of $c$.

Requiring the synchrotron cooling time scales $t'_{\rm syn}(E'_{\pi, \mu})$ to exceed the decay time scales $t'_{\rm dec}(E'_{\pi,\mu}) = (E'_{\pi,\mu}/m_{\pi, \mu})\tau_{\pi,\mu}$, where $m_{\pi,\mu}$ and $\tau_{\pi,\mu}$ are the relevant masses and lifetimes, we obtain two additional constraints:    
\begin{align}
R_{17} \Gamma^2 \gtrsim \xi_{\pi,\mu} \frac{E_{\nu, \rm ob} \hat z}{\rm PeV} \Big( L^{\rm iso}_{\rm R, 45} \frac{0.03}{\epsilon_{\rm R}} \cdot \frac{0.1}{\epsilon_e}\cdot \frac{\epsilon_B}{0.1} \Big)^{1/2},\label{eq:cooling}
\end{align} 
where $\xi_\pi \approx 1.1\times 10^{-3}$, $\xi_\mu \approx 10^{-2}$, and we have taken $E'_\pi \approx 2E'_\mu \approx 0.2E'_p$.

\subsection{Non-radiation-mediated shock}  
In addition, for efficient particle acceleration, the shock should not be radiation-mediated \citep{Murase2013}, which requires
\begin{align}
R_{17} \Gamma^3 \gtrsim 5 \times 10^{-4}~ L_{\rm R, 45}^{\rm iso} \frac{0.03}{\epsilon_{\rm R}} \cdot \frac{0.1}{\epsilon_e} \label{eq:rad} 
\end{align} 
for shocks similar to internal shocks. The above constraint is the same as Equation~(5) in \citet{Murase2013} for 
a typical relative Lorentz factor $\Gamma_{\rm rel}=10^{0.5}$ between fast and merged shells. 

\subsection{Constraints and implications} 
Taking $L^{\rm iso}_{\rm R}=10^{45}$ erg/s, $\epsilon_{\rm R}=0.03$, $\epsilon_e=0.1$, $\epsilon_B=0.1$, $\hat z=1$,
and $E_{\nu, \rm ob} = 2$ PeV, we show all of the above constraints in Figure \ref{fig:RGamma}. It can be seen that
efficient production of PeV neutrinos from late GRB bumps only occurs in a small region of the $R$--$\Gamma$ space,
e.g., $R\sim 2 \times 10^{15}$--$10^{17}$~cm at $\Gamma\sim 1$ and $R\sim 8 \times 10^{13}$ to $2\times 10^{14}$~cm at $\Gamma\sim 5$.
   
We now compare several models with the constraints in Figure \ref{fig:RGamma}. Consider an adiabatic external blast wave with a total energy $E_0$ propagating in the interstellar medium (ISM) with a uniform density $n_0$. Its radius evolves as $R(t_{\rm ob}) \approx 4 \Gamma^2(t_{\rm ob}) ct_{\rm ob}/\hat z$ with $\Gamma(t_{\rm ob}) \approx 7(E_{0,53}/n_0)^{1/8} (t_{\rm ob, day}/\hat z)^{-3/8}$ \citep{Razzaque2013}. The $R$--$\Gamma$
relation for typical GRBs with $10^{-2}<E_{0,53}/n_0<10^{2}$ at $t_{\rm ob} = 1$ day is shown as the red crosses in Figure \ref{fig:RGamma}
and corresponds to $f_{p\gamma} \sim 10^{-6}$--$10^{-2}$. The resulting neutrino fluence is further suppressed because the associated afterglow is much less luminous than the bumps. Similarly, models involving a two-component jet or refreshed shock are not efficient for making HE neutrinos associated with the late bumps, as they have similar or even larger values of $R$ and $\Gamma$ compared to the external blast wave.

Models invoking density bumps can
generally have $f_{p\gamma}\simeq 1$ with low $\Gamma$ \citep[see, e.g.,][]{Lazzati2002,Dai2003}.
However, for the energy deposited in the shocked ISM to account for the observed brightness of the late bump, 
the ISM density needs to be high enough. For a blast wave in an ISM with a constant density $n_0$, the internal energy of the shocked ISM in the stellar rest frame is $E_{\rm ISM} \sim \frac{4}{3}\pi R^3 n_0 m_p c^2 \Gamma(\Gamma-1)$ \citep{Kumar2014}. Considering a simple case where the EM bump is mainly emitted from the shocked ISM, i.e., neglecting the contributions from the reverse shock, the energy radiated in the $R$ band is $\sim E_{\rm ISM}\epsilon_e \epsilon_{\rm R}$, which is required to match the observed energy $E_{\rm R}^{\rm iso} \sim T_{\rm dur} L_{\rm R}^{\rm iso} \sim 10^{50}$ erg, with $T_{\rm dur}$ being the duration of the bump. Considering $R_{17} \Gamma^4 \lesssim 1$ at $t_{\rm ob} \sim$ 1 day for efficient production of $\sim$ PeV neutrinos [see Equation~(\ref{eq:fpr})], we obtain that $n_0$ should be as high as $\sim 5 \times 10^{3}~{\rm cm^{-3}}~\Gamma^{11}(\Gamma-1)^{-1} E_{\rm R, 50}^{\rm iso} (0.1/\epsilon_e)(0.03/\epsilon_{\rm R})$. Due to the high power dependence on $\Gamma$, density bump models only work for nonrelativistic or mildly relativistic shocks with $\Gamma \lesssim 2$. 
Another possible scenario is the interaction between a slightly later jet with the cocoon driven by the prompt GRB jet, where the bump/flare at late times can arise from nonrelativistic or mildly relativistic shocks \citep{Shen2010}. 
Note that any such outflow with $\Gamma\lesssim 2$ launched at a time $\ll 1$~day
would reach $R\sim 10^{15}$--$10^{16}$~cm at $t_{\rm ob}\simeq 1$~day (see the top dashed curve in Figure \ref{fig:RGamma}), consistent with the allowed region shown in Figure \ref{fig:RGamma}.

\begin{figure}
\centering
\includegraphics[width=7.0cm]{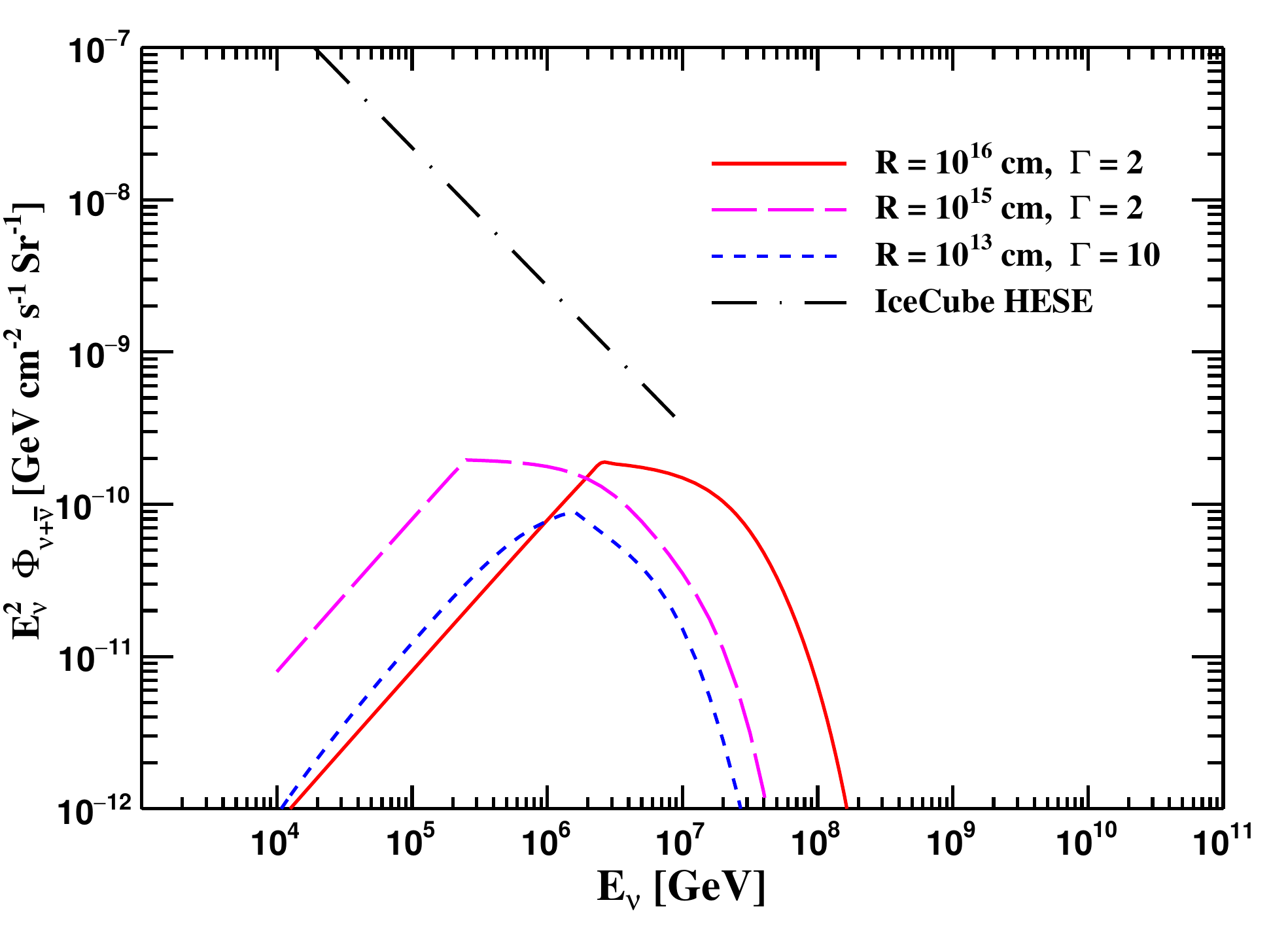}%
\caption{Diffuse neutrino background per flavor from late bumps for ($R$/cm, $\Gamma$) $=$ ($10^{16}$, 2), ($10^{15}$, 2), and ($10^{13}$, 10). The observed flux at IceCube fitted to a single power law \citep{HESE2019} is shown for comparison. See the text for details.}   
\label{fig:Flux} 
\end{figure}       
               
\section{Diffuse flux and events expected at IceCube}

We now estimate the expected flux of HE neutrinos produced by GRB sources with late bumps considering the above constraints.
We take $F_p\equiv dN_p/dE_p = A_p E_p^{-2} e^{-E_p/E_p^{\max}}$ as the cosmic-ray spectrum for a typical GRB.
Assuming most of the shock energy goes to accelerating protons, we have $\int F_p E_p dE_p \approx E_{\rm bump}/\epsilon_e$, where $E_{\rm bump} \sim  E_{\rm R}^{\rm iso}/\epsilon_{\rm R}$ is the isotropic energy emitted in all EM bands from the bump. We take $\int F_p E_p dE_p \sim A_p\ln(E_p^{\rm max}/E_p^{\rm min}) \sim 15 A_p$ for $E_p^{\rm min} \sim 10$ GeV and $E_p^{\rm max} \sim 10^{7-10}$ GeV.
The diffuse neutrino background flux per flavor from the late bumps can then be estimated as \citep{Murase2006,Murase2013}
\begin{align}\label{eq:flux}
E_\nu^2 \Phi_{\nu+\bar\nu} \sim & \frac{c}{4\pi H_0} f_{\rm bump} F_p E_p^2  \frac{\min[1,f_{p\gamma}]}{ 2\times 4 }R_{\rm GRB}(0) f_z f_{\rm sup} \nonumber \\
 \approx & 2\times 10^{-10}~{\rm GeV ~cm^{-2}~ s^{-1}~ sr^{-1}}  \min[1, f_{p\gamma}] \nonumber \\
 & \times
\left(\frac{R_{\rm GRB}(0)}{2 ~{\rm Gpc^{-3} ~yr^{-1}}}\right) \left(\frac{f_z}{3}\right) \left(\frac{f_{\rm bump}}{0.2}\right) f_{\rm sup} \nonumber \\ 
& \times \left(\frac{0.1}{\epsilon_e}\right)\left(\frac{0.03}{\epsilon_{R}}\right) \left(\frac{E
_{\rm R}^{\rm iso}}{10^{50} ~{\rm erg}}\right),
\end{align} 
where $H_0 \approx 70~\rm{km~s^{-1}~Mpc^{-1}}$ is the Hubble constant, the factor $1/2$ reflects that only half of the $p\gamma$ reactions produce $\pi^\pm$, the factor $1/4$ accounts for the average ratio of $E_\nu/E_\pi$ in $\pi^\pm$ decay, $R_{\rm GRB}(0)$ is the local GRB rate, and $f_z \sim 3$ is the evolution factor \citep{waxman1998}. In the above equation, $f_{\rm sup}$ is the suppression factor due to secondary pion and muon cooling, and can be approximated as \citep{Razzaque2004,Razzaque2005}
\begin{align}
 f_{\rm sup} \sim & \frac{ t'^{-1}_{\rm dec, \pi}(E'_\pi)}{ t'^{-1}_{\rm dec, \pi}(E'_\pi) + t'^{-1}_{\rm syn, \pi}(E'_\pi)} \nonumber \\ 
 & \times
 \left( \frac{1}{3} + \frac{2}{3} \cdot \frac{ t'^{-1}_{\rm dec, \mu}(E'_\mu) }{  t'^{-1}_{\rm dec, \mu}(E'_\mu) + t'^{-1}_{\rm syn, \mu}(E'_\mu)} \right),      
\end{align}
with $E'_\pi \approx 2 E'_\mu \approx 0.2 E'_p$, and $f_{\rm bump}$ is the fraction of GRBs with late bumps at $t_{\rm ob}\sim 1$ day. \citet{Li2012} and \citet{Liang2013} collected a total of 146 GRBs from 1997 February to 2011 November, which had well-sampled optical light curves extending up to $10^3$--$10^7$ s after the burst. \footnote{The GRB sample examined in \citet{Liang2013} is different from the sample used here for the correlation analysis, as they cover different periods of observation. An updated analysis of the long-term afterglow properties of a more recent and complete GRB sample would be desirable.} They found about 10 GRBs with late optical bumps at $\sim$1 day. 
Therefore, we expect $f_{\rm bump}\sim 0.1$.
As nonrelativistic outflows have wider opening angles than the prompt jet, there could be orphan optical bumps, which, however, are difficult to observe due to the lack of triggers by the prompt $\gamma$-rays. So $f_{\rm bump}$ could be as high as $\sim$1.

Taking $L^{\rm iso}_{R}=10^{45}$ erg/s, $E^{\rm iso}_{R}=10^{50}$ erg, $\epsilon_{R}=0.03$, $\epsilon_e=0.1$, $\epsilon_B=0.1$, $R_{\rm GRB}(0)=2~{\rm Gpc^{-3}~yr^{-1}}$ \citep{Wanderman10,Lan2019}, $f_{\rm bump} = 0.2$, and $f_z=3$, we compute the diffuse flux for a broad range of $E_\nu$ considering the details of
the $p\gamma$ reactions (see Appendix~\ref{sec:pr}) and show in Figure \ref{fig:Flux} 
the results for ($R$/cm, $\Gamma$) $=$ ($10^{16}$, 2), ($10^{15}$, 2), and ($10^{13}$, 10). 
The linear rise in the flux at lower energy is due to the increase of $f_{p\gamma}$ up to the peak with $f_{p\gamma} = 1$. 
The decline of the flux is due to meson cooling and at very high energies, to the lack of protons above $E_p^{\rm max}$. For a fixed $\Gamma$, increase in $R$ shifts the neutrino flux to higher energies in accord with
$E_\nu^{\rm max}\propto(R\Gamma^3)^{1/2}$ [Equation~(\ref{eq:acc})],
$f_{p\gamma}\propto E_\nu/(R\Gamma^4)$ [Equation~(\ref{eq:fpr0})],
and the energy $E_\nu \propto R \Gamma^2$ for significant meson cooling [Equation~(\ref{eq:cooling})].
For $(R/{\rm cm}, \Gamma)=(10^{16}, 2)$ and $(10^{13}, 10)$, $f_{p\gamma} \propto R^{-1}\Gamma^{-4}$ are similar but
meson cooling takes effect at a lower energy for the latter, with 
the corresponding flux at $\gtrsim 1$~PeV more suppressed.
For comparison, the observed flux per flavor from the HE Starting Events (HESEs) at IceCube in 7.5 yr,  
$E_\nu^2\Phi_{\nu+\bar\nu} \approx 2.2 \times 10^{-8}(E_\nu/100~{\rm TeV})^{-0.91}~{\rm GeV~cm^{-2}~s^{-1}~Sr^{-1}}$ \citep{HESE2019}, is 
also shown in Figure \ref{fig:Flux}. Using the effective area from \url{http://icecube.wisc.edu/science/data/HE-nu-2010-2012}, we estimate $\sim$ 0.5, 1.0, and 0.3 events with $E_\nu\sim 0.1$--10 PeV at IceCube in 6 yr, for ($R$/cm, $\Gamma$) $=$ ($10^{16}$, 2), ($10^{15}$, 2), and ($10^{13}$, 10), respectively, which is
broadly consistent with our correlation analysis in view of uncertainties in Equation~\eqref{eq:flux}.

\section{Summary and discussions}

We have shown that the IceCube data allow 
up to $\sim$10\% of the observed events to be associated with late-time emission of GRBs at $\sim$1~day.
If such delayed neutrinos have the same origin as the observed late-time bumps in GRBs, strong constraints on viable mechanisms for these bumps can be derived.
In particular, the shocked outflow producing the bump would have to be nonrelativistic or mildly relativistic.
So models involving external blast waves in the ISM, such as refreshed shocks and two-component jets, would be disfavored. 

For most of the possibly correlated pairs in the data that we have analyzed, the IceCube events are shower events with direction uncertainties of $\sigma_\nu\sim 10^\circ$. The GRBs observed by the $Fermi$ Gamma-ray Burst Monitor (GBM) also have relatively large direction uncertainties. Both factors severely limit the significance of the correlation. In contrast, using $\sigma_\nu \sim 1^\circ$ for track events and $\sigma_{\rm grb}\ll 1^\circ$
for GRBs, we estimate that two such correlated pairs correspond to a significance of $\sim$3$\sigma$. As a hypothetical example,
this level of significance would have been achieved if the two track events (ID-63 and ID-23) listed in Table~\ref{tab:pair} both had been separated by $\lesssim 1^\circ$ from the corresponding GRB counterparts (GRB141207A and GRB120121C) and if GRB120121C had also been precisely localized.

If more than several correlated neutrino events were observed in the future, then the baryon loading, the total energy budget, and the occurrence rate of the bumps could be better probed. Furthermore, the magnetic origin for late bumps would be disfavored because
the associated shocks are much weaker and HE neutrino production is suppressed \citep{Murase2006}. In principle, the nondetection of correlation might also be used to constrain the parameters $R$, $\Gamma$, $f_{\rm bump}$, and $\epsilon_e$, etc. For example, taking the simple case with $f_{p\gamma} \approx 1$ for making PeV neutrinos [see the case of ($R$/cm, $\Gamma$)=($10^{15}$, 2) shown in Figure~\ref{fig:Flux}], current data require $(f_{\rm bump}/0.2) (0.1/\epsilon_e)(0.03/\epsilon_{\rm R}) \lesssim 8$. This constraint, however, will be greatly relaxed if the shocks are highly relativistic with $f_{p\gamma} \ll 1$. No strong constraints can be put on $R$ and $\Gamma$ from the null result at present. Such constraints could be possible if more IceCube events, especially track events, are accumulated in the future, and if more GRBs are well localized.
When such data become available, detailed studies could result in either a true physical correlation or stronger constraints on our proposed scenario.     

We have only considered the potential HE neutrino signals from late GRB bumps and the strong constraints on the associated shocks.
For such late GRB neutrinos from nonrelativistic or mildly relativistic shocks, contributions from the $pp$ process are severely limited \citep{Murase2013,Senno2016} and can be ignored. In addition, the constraints from diffuse $\gamma$-rays are easily satisfied due to a large opacity for $\gamma$-rays \citep{Murase2016}. In order to identify the detailed signatures for the connection between HE neutrinos and GRB bumps,
further studies of the EM signals in the X-ray/optical bumps need to be pursued, under consideration of the constraints derived here.
The delayed HE neutrinos could be expected from both long and short GRBs. With a nearby short GRB from a binary neutron star merger, such neutrinos would add to the multi-messenger observations in gravitational waves, broadband EM radiation, and HE neutrinos on different time scales \citep{Kimura2017}.

\acknowledgments 
This work was supported in part by the Ministry of Science and Technology, Taiwan under grants No. 107-2119-M-001-038 and No. 108-2112-M-001-010, the Physics Division of the National Center for Theoretical Sciences (G.G., M.R.W.), and the US Department of Energy [DE-FG02-87ER40328 (Y.Z.Q.)].



\appendix
\twocolumngrid

\section{Calculation of $\MakeLowercase{f_{p\gamma}}$} 
\label{sec:pr}

In the comoving frame of the shocked outflow, the time scale for proton cooling due to $p\gamma$ reactions can be estimated from 
\begin{align}
t^{\prime-1}_{p\gamma}(E_p^\prime) 
= &c\int d\epsilon^\prime d\Big[{\cos\theta^\prime\over 2}\Big] \nonumber \\ 
& \times \kappa(\epsilon^{\prime\prime})\sigma_{p\gamma}(\epsilon^{\prime\prime}) {dn^\prime(\epsilon^\prime) \over d\epsilon^\prime}(1-\cos\theta^\prime),  \label{eq:tpg}
\end{align}  
where $E_p^\prime$ and $\epsilon^\prime$ are the energies of the proton and the photon, respectively, $\theta^\prime$ is their intersection angle,
$dn^\prime(\epsilon^\prime)/d\epsilon^\prime$ is the energy-differential density of photons,
$\epsilon^{\prime\prime} \equiv (1-\cos\theta^\prime)E_p^\prime \epsilon^\prime/(m_pc^2)$ is the photon energy in the proton rest frame, 
$\sigma_{p\gamma}$ is the cross section, and $\kappa$ is the energy fraction transferred to pions (inelasticity). The factor 1/2 accounts for the
approximately isotropic distribution of photons in the comoving frame.
 
We use
\begin{align}
\kappa \sigma_{p\gamma}(\epsilon'')\approx 
\left\{ 
\begin{array}{cc}
  \kappa_1 \sigma_{\Delta}(\epsilon^{\prime\prime}),  & 0.15 \le \epsilon'' < 0.5~\mathrm{GeV},  \\
  \kappa_1 \times 2.0 \times 10^{-28}\mathrm{~cm^2}, & 0.5 \le \epsilon''< 1.2~\mathrm{GeV},  \\
  \kappa_2 \times 1.2 \times 10^{-28} \mathrm{~cm^2},  &  \epsilon'' \ge 1.2~\mathrm{GeV},  
\end{array}
\right.   
\end{align}
where $\sigma_\Delta(\epsilon'') = \big(\frac{s}{\epsilon''}\big)^2 \frac{\sigma_0 \Gamma_\Delta^2}{(s-M_\Delta^2)^2+\Gamma_\Delta^2 s}$ is the cross section for the $\Delta$-resonance with $s=m_p^2+2m_p\epsilon''$ being the center-of-mass energy squared, $M_\Delta = 1.23$ GeV, $\Gamma_\Delta=0.11$ GeV, and $\sigma_0 \approx 0.3 \times 10^{-28}~{\rm cm}^2$. We take $\kappa_1 \sim 0.2$ for the resonance channels and $\kappa_2 \sim 0.5$ for the multi-pion production channels \citep{mucke2000}. 

\begin{figure} 
\centering
\includegraphics[width=7.5cm]{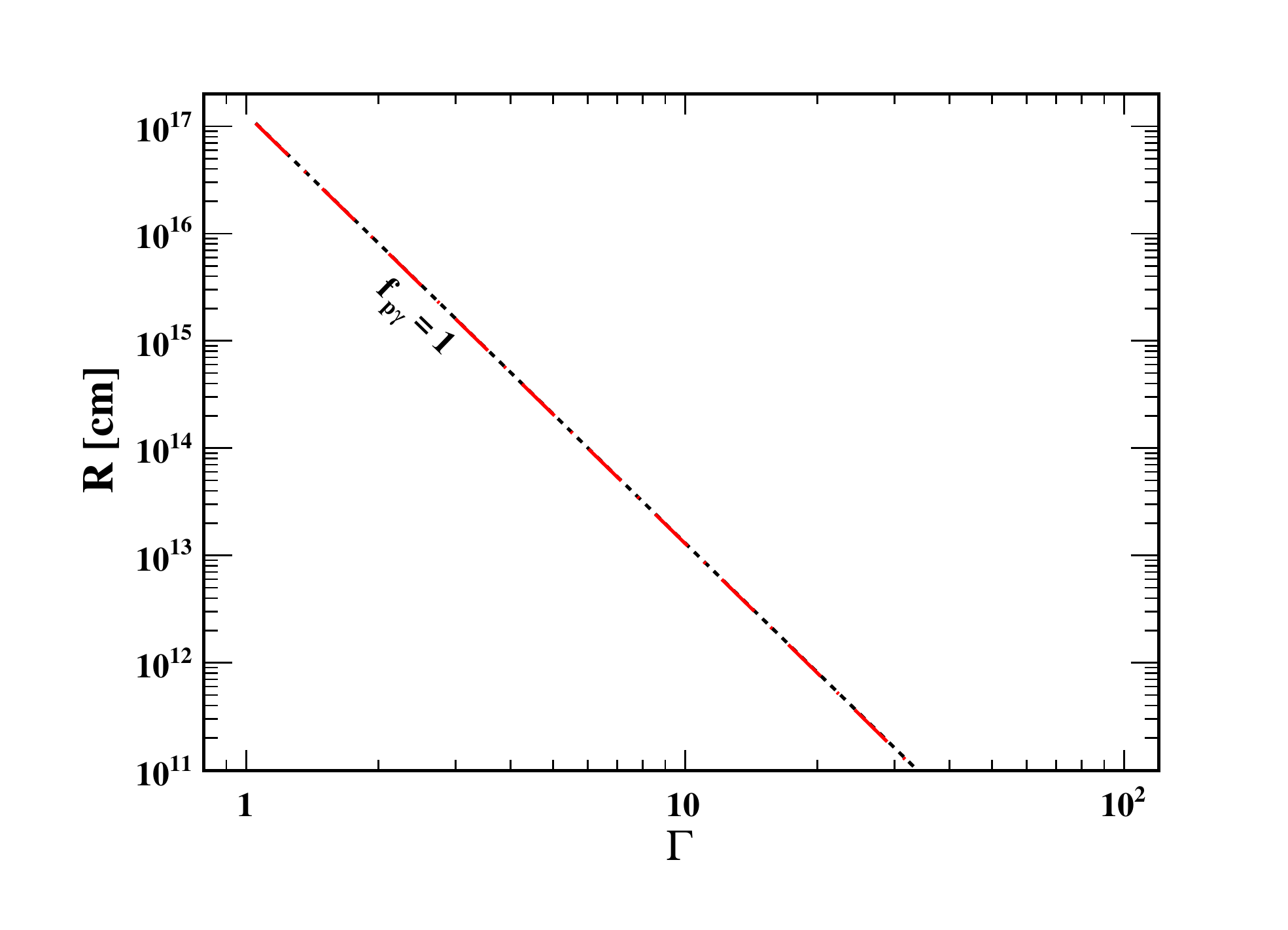}%
\caption{Comparison of $R$--$\Gamma$ contours for $f_{p\gamma}=1$ based on detailed integration (red dashed curve) and
analytical approximation (black dotted curve). See the text for details.} 
\label{fig:RGamma_num} 
\end{figure}

Taking $L^{\rm iso}_{\rm R}=10^{45}$ erg/s and $E_{\nu, {\rm ob}}\hat z=2$ PeV, we show in Figure \ref{fig:RGamma_num} the $R$--$\Gamma$ contour for $f_{p\gamma}=1$ based on the integral in Equation~(\ref{eq:tpg}). This result is indistinguishable from the analytical expression in the $\Delta$-resonance approximation [see Equation~(\ref{eq:fpr0}) of the main text].


\begin{thebibliography}{}
\expandafter\ifx\csname natexlab\endcsname\relax\def\natexlab#1{#1}\fi
\providecommand{\url}[1]{\href{#1}{#1}}
\providecommand{\dodoi}[1]{doi:~\href{http://doi.org/#1}{\nolinkurl{#1}}}
\providecommand{\doeprint}[1]{\href{http://ascl.net/#1}{\nolinkurl{http://ascl.net/#1}}}
\providecommand{\doarXiv}[1]{\href{https://arxiv.org/abs/#1}{\nolinkurl{https://arxiv.org/abs/#1}}}


\bibitem[{Aartsen {et~al.}(2013)}]{evidence}
Aartsen, M.~G., {et~al.} 2013, Sci, 342, 1242856

\bibitem[{Aartsen {et~al.}(2014{\natexlab{a}})}]{3years}
---. 2014{\natexlab{a}}, PhRvL, 113, 101101

\bibitem[{Aartsen {et~al.}(2014{\natexlab{b}})}]{Gen2}
---. 2014{\natexlab{b}}, arXiv:1412.5106

\bibitem[{Aartsen {et~al.}(2015{\natexlab{a}})}]{4years}
---. 2015{\natexlab{a}}, arXiv:1510.05223

\bibitem[{Aartsen {et~al.}(2015{\natexlab{b}})}]{grb2014}
---. 2015{\natexlab{b}}, ApJL, 805, L5

\bibitem[{Aartsen {et~al.}(2016{\natexlab{a}})}]{grb2016}
---. 2016{\natexlab{a}}, ApJ, 824, 115

\bibitem[{Aartsen {et~al.}(2016{\natexlab{b}})}]{HE1}
---. 2016{\natexlab{b}}, PhRvL, 117, 241101

\bibitem[{Aartsen {et~al.}(2016{\natexlab{c}})}]{HE2}
---. 2016{\natexlab{c}}, ApJ, 833, 3


\bibitem[{Aartsen {et~al.}(2017a)}]{Aartsen:2017mau}
---. 2017{\natexlab{a}}, arXiv:1710.01191

\bibitem[{Aartsen {et~al.}(2017{\natexlab{b}})}]{Aartsen2017_grb}
---. 2017{\natexlab{b}}, ApJ, 843, 112

\bibitem[{Aartsen {et~al.}(2018)}]{Aartsen2018}
---. 2018, Sci, 361, eaat1378

\bibitem[{Abbasi {et~al.}(2010)}]{grb2010}
Abbasi, R., {et~al.} 2010, ApJ, 710, 346

\bibitem[{Abbasi {et~al.}(2011)}]{grb2011}
---. 2011, PhRvL, 106, 141101

\bibitem[{Ackermann {et~al.}(2018)Ackermann, Fermi-LAT, MAGIC, AGILE, Kanata,
  Kiso, Kapteyn, Subaru, \& VERITAS}]{Ackermann2018}
Ackermann, M., {et~al.} 2018, Sci, 361, 147


\bibitem[{Adrian-Martinez {el~al.}(2016)}]{KM3NeT} 
  Adrian-Martinez, S., et al. 2016,
  JPhG, 43, 084001

\bibitem[{Berger {et~al.}(2003)}]{Berger2003}
Berger, E., {et~al.} 2003, Natur, 426, 154

\bibitem[{Braun {et~al.}(2008)Braun, Dumm, De~Palma, Finley, Karle, \&
  Montaruli}]{braun2008}
Braun, J., Dumm, J., De~Palma, F., {et~al.} 2008, APh, 29, 299

\bibitem[{Burrows {et~al.}(2005)}]{Burrows:2005}
Burrows, D.~N., {et~al.} 2005, Sci, 309, 1833

\bibitem[{Casey(2015)}]{Casey2015}
Casey, J. 2015, PhD thesis, Georgia Institute of Technology

\bibitem[{Connaughton {et~al.}(2015)}]{Connaughton2014}
Connaughton, V., {et~al.} 2015, ApJS, 216, 32

\bibitem[{Dai \& Lu(2001)}]{Dai2000}
Dai, Z.~G., \& Lu, T. 2001, ApJ, 551, 249

\bibitem[{Dai \& Wu(2003)}]{Dai2003}
Dai, Z.~G., \& Wu, X.~F. 2003, ApJL, 591, L21

\bibitem[{Dermer(2002)}]{Dermer2000}
Dermer, C.~D. 2002, ApJ, 574, 65


\bibitem[{Huber(2019)}]{Huber2019} 
M.~Huber 2019, arXiv:1908.08458 

\bibitem[{Kann {et~al.}(2010)}]{Kann2010}
Kann, D.~A., {et~al.} 2010, ApJ, 720, 1513

\bibitem[{Kimura {et~al.}(2017)}]{Kimura2017}  
  Kimura, S.~S., Murase K., M\'esz\'aros P., \& Kiuchi, K. 2017,
  ApJL, 848, L4

\bibitem[{Kobayashi \& Zhang(2003)}]{Kobayashi2003}
Kobayashi, S., \& Zhang, B. 2003, ApJ, 597, 455

\bibitem[{Kopper(2018)}]{Kopper2017}
Kopper, C. 2018, ICRC (Busan), 35, 981

\bibitem[{Kumar \& Piran(1999)}]{Kumar1999}
Kumar, P., \& Piran, T. 1999, ApJ, 523, 286

\bibitem[{Kumar \& Zhang(2014)}]{Kumar2014}
Kumar, P., \& Zhang, B. 2014, PhR, 561, 1

\bibitem[{Lan {et~al.}(2019)Lan, Zeng, Wei, \& Wu}]{Lan2019}   
Lan, G. X., Zeng, H. D., Wei, J. J., \& Wu, X. F. 2019,
MNRAS, 488, 4607  

\bibitem[{Lazzati {et~al.}(2002)Lazzati, Rossi, Covino, Ghisellini, \&
  Malesani}]{Lazzati2002}
Lazzati, D., Rossi, E., Covino, S., Ghisellini, G., \& Malesani, D. 2002,
  A\&A, 396, L5
  
\bibitem[{{Li} {et~al.}(2012)}]{Li2012}
{Li}, L., {et~al.} 2012, ApJ, 758, 27

\bibitem[{Li {et~al.}(2002)Li, Dai, \& Lu}]{Li2002}
Li, Z., Dai, Z.~G., \& Lu, T. 2002, A\&A, 396, 303

\bibitem[{{Liang} {et~al.}(2013)}]{Liang2013}
{Liang}, E.-W., {et~al.} 2013, \apj, 774, 13

\bibitem[{{Margutti} {et~al.}(2010)}]{Margutti2010}
{Margutti}, R., {et~al.} 2010, MNRAS, 402, 46

\bibitem[{Melandri {et~al.}(2014)}]{Melandri:2014}
Melandri, A., {et~al.} 2014, A\&A, 572, A55

\bibitem[{M\'esz\'aros(2006)}]{Meszaros2006}
M\'esz\'aros, P. 2006, RPPh, 69, 2259

\bibitem[{M\'esz\'aros \& Waxman(2001)}]{Meszaros2001}
M\'esz\'aros, P., \& Waxman, E. 2001, PhRvL, 87, 171102

\bibitem[{M{\"u}cke {et~al.}(2000)M{\"u}cke, Engel, Rachen, Protheroe, \&
  Stanev}]{mucke2000}
M{\"u}cke, A., Engel, R., Rachen, J.~P., Protheroe, R.~J., \& Stanev, T. 2000,
CoPhC, 124, 290

\bibitem[{Murase(2007)}]{Murase2007}
Murase, K. 2007, PhRvD, 76, 123001

\bibitem[{Murase {et~al.}(2016)Murase, Guetta, \& Ahlers}]{Murase2016}
Murase, K., Guetta, D., \& Ahlers, M. 2016, PhRvL, 116, 071101

\bibitem[{Murase \& Ioka(2013)}]{Murase2013}
Murase, K., \& Ioka, K. 2013, PhRvL, 111, 121102

\bibitem[{Murase \& Nagataki(2006)}]{Murase2006}
Murase, K., \& Nagataki, S. 2006, PhRvL, 97, 051101



\bibitem[{Piran(2005)}]{piran2005}
Piran, T. 2005, RvMP, 76, 1143

\bibitem[{Rachen \& Meszaros(1998)}]{Rachen:1998}
Rachen, J.~P., \& Meszaros, P. 1998, PhRvD, 58, 123005

\bibitem[{Razzaque(2013)}]{Razzaque2013}
Razzaque, S. 2013, PhRvD, 88, 103003

\bibitem[{Razzaque {et~al.}(2003)Razzaque, M\'esz\'aros, \&
  Waxman}]{Razzaque2003}
Razzaque, S., M\'esz\'aros, P., \& Waxman, E. 2003, PhRvD, 68, 083001

\bibitem[{Razzaque {et~al.}(2004)Razzaque, M\'esz\'aros, \&
  Waxman}]{Razzaque2004}
---. 2004, PhRvL, 93, 181101

\bibitem[{Razzaque {et~al.}(2005)Razzaque, Meszaros, \& Waxman}]{Razzaque2005}
Razzaque, S., Meszaros, P., \& Waxman, E. 2005, MPLA, 20, 2351

\bibitem[{Razzaque \& Yang(2015)}]{Razzaque2014}
Razzaque, S., \& Yang, L. 2015, PhRvD, 91, 043003

\bibitem[{Rees \& M\'esz\'aros(1998)}]{Rees1997}
Rees, M.~J., \& M\'esz\'aros, P. 1998, ApJL, 496, L1

\bibitem[{Sari \& M\'esz\'aros(2000)}]{Sari2000}
Sari, R., \& M\'esz\'aros, P. 2000, ApJL, 535, L33

\bibitem[{Senno {et~al.}(2016)Senno, Murase, \& M\'esz\'aros}]{Senno2016}
Senno, N., Murase, K., \& M\'esz\'aros, P. 2016, PhRvD, 93, 083003

\bibitem[{{Shen} {et~al.}(2010){Shen}, {Kumar}, \& {Piran}}]{Shen2010}
{Shen}, R., {Kumar}, P., \& {Piran}, T. 2010, MNRAS, 403, 229

\bibitem[{Stachurska(2019)}]{HESE2019}
Stachurska, J. 2019, EPJWC, 207, 02005

\bibitem[{Thomas {et~al.}(2017)Thomas, Moharana, \& Razzaque}]{Thomas2017}
Thomas, J.~K., Moharana, R., \& Razzaque, S. 2017, PhRvD, 96, 103004

\bibitem[{Wanderman \& Piran(2010)}]{Wanderman10} 
Wanderman, D., \& Piran, T. 2010, MNRAS, 406, 1944

\bibitem[{Waxman(1995)}]{Waxman95}
Waxman, E. 1995, PhRvL, 75, 386

\bibitem[{Waxman \& Bahcall(1997)}]{waxman1997}
Waxman, E., \& Bahcall, J. 1997, PhRvL, 78, 2292

\bibitem[{Waxman \& Bahcall(1998)}]{waxman1998}
---. 1998, PhRvD, 59, 023002

\bibitem[{Waxman \& Bahcall(2000)}]{Waxman1999}
Waxman, E., \& Bahcall, J.~N. 2000, ApJ, 541, 707

\bibitem[{Zaninoni {et~al.}(2013)Zaninoni, Bernardini, Margutti, Oates, \&
  Chincarini}]{Zaninoni2013}
Zaninoni, E., Bernardini, M.~G., Margutti, R., Oates, S., \& Chincarini, G.
  2013, A\&A, 557, A12

\bibitem[{Zhang {et~al.}(2006)Zhang, Fan, Dyks, Kobayashi, Meszaros, Burrows,
  Nousek, \& Gehrels}]{Zhang:2005}
Zhang, B., Fan, Y.~Z., Dyks, J., {et~al.} 2006, ApJ, 642, 354

\bibitem[{Zhang {et~al.}(2003)Zhang, Kobayashi, \& M\'esz\'aros}]{Zhang2003}
Zhang, B., Kobayashi, S., \& M\'esz\'aros, P. 2003, ApJ, 595, 950

\end{thebibliography}

            
\end{document}